\documentstyle[11pt,newpasp,twoside]{article}
\def\edcomment#1{\iffalse\marginpar{\raggedright\sl#1\/}\else\relax\fi}
\reversemarginpar

\begin{document}
\title{PHL 1811: The Local Prototype of the Lineless High-z SDSS QSOs}
\author{Karen M. Leighly}
\affil{The University of Oklahoma, Department of Physics and
Astronomy, 440 W. Brooks St., Norman, OK 73019}
\author{Jules P. Halpern}
\affil{Department of Astronomy, Columbia University, 550 W.\ 120th St., New York,
NY 10025-6601}
\author{Edward B. Jenkins}
\affil{Princeton University Observatory, Princeton, NJ, 08544-1001}

\begin{abstract}
In the SDSS, several unusual QSOs have been discovered that have very
blue rest-frame UV spectra but no discernible emission lines.  Their
UV spectra strongly resemble that of the newly discovered quasar PHL
1811 ($z=0.192$; $\rm M_{V}=-25.9$).

With magnitudes of $B = 14.4$ and $R = 14.1$, PHL 1811 is the second
brightest quasar known with $z>0.1$ after 3C 273.  Optically it is
classified as a Narrow-line Seyfert 1 galaxy (NLS1).  Objects of this
class are generally strong soft X-ray emitters, but a BeppoSAX
observation of PHL 1811 showed that it is anomalously X-ray weak.  The
inferred $\alpha_{ox}$ was 1.9--2.1, much steeper than the nominal value of
1.6 for quasars of this optical luminosity, and comparable to the
X-ray weakest quasars.  Follow-up {\it Chandra} observations reveal a
variable, unabsorbed X-ray spectrum and confirm that it is
intrinsically X-ray weak.

{\it HST} STIS spectra of PHL 1811 reveal a very blue continuum with
little evidence for absorption or scattering intrinsic to the quasar.
High-ionization lines are very weak; \ion{C}{IV} has an equivalent
width of only $\sim 5$\AA\/.  Neither forbidden nor semiforbidden
emission lines are detected. \ion{Fe}{II} is the dominant line
emission in the UV.  High metallicity is implied by the large
\ion{Fe}{II} to \ion{Mg}{II} ratio and relatively strong \ion{N}{V}.
Low-ionization emission lines of \ion{Al}{III}, \ion{Na}{I} D, and
\ion{Ca}{II} H \& K are present, implying high optical depth.

We demonstrate that the emission-line properties of PHL 1811 are a
direct consequence of the UV-peaked continuum and weak X-ray emission.
We propose that these properties are a consequence of high accretion
rate, which powers the UV emission from an optically thick accretion
disk, while suppressing the formation of a hot corona.  This is an
extreme case of the same mechanism which is thought to be responsible
for luminous NLS1s.  Based on the similarity between PHL 1811 and the
lineless SDSS quasars, we propose that the lineless quasars discovered
in the SDSS are the high-z counterparts of local high-luminosity
NLS1s.

\end{abstract}

\section{Introduction}

In the Sloan Digital Sky Survey, a number of high redshift quasars
have been discovered that are remarkable for their lack of emission
lines.  A detailed study of one of these objects,
SDSSp~J153259.96$-$003944.1 (hereafter referred to as SDSS~J1532$-$0039)
was presented by Fan et al.\ 1999.  The optical spectrum, redward of
6800 \AA\/, is a featureless blue power-law continuum; there are no
emission lines.  Blueward of 6800 \AA\/, there are features due to
Ly$\alpha$ absorption that lead to a redshift estimate of $4.52$.
Featureless continua are frequently found in Bl Lac objects. However, 
SDSS~J1532$-$0039 does not appear to have other characteristic properties
of Bl~Lac objects: it was not detected in a deep radio observation,
doesn't vary in the optical, and was not found to be optically
polarized.  Follow-up observations with {\it Chandra} failed to detect the
quasar, indicating an $\alpha_{ox}>1.74$ \footnote{$\alpha_{ox}$ is
defined as the point-to-point slope between 2500\AA\/ and 2 keV.}.
This steep $\alpha_{ox}$ makes it moderately X-ray weak (Vignali et
al.\ 2001).   

PHL 1811 is a luminous narrow-line quasar that was rediscovered in the
VLA FIRST survey (Leighly et al.\ 2000).  It is a very bright ($B=14.4$,
$R=14.1$), nearby ($z=0.192$) object, well away from the Galactic plane,
so it was surprising that it was not detected in the ROSAT All Sky
Survey. Two {\it Chandra} observations show that it is X-ray weak,
with inferred $\alpha_{ox}$ between $1.9$ and $2.1$, much steeper than
for a typical quasar of this luminosity ($1.6$; Wilkes et al.\ 1994).
The {\it HST} spectrum is unusual in that it is dominated by
\ion{Fe}{II}, and there are no prominent broad emission lines.  We
thought that the X-ray weakness and the lack of broad emission lines
suggest a similarity to SDSS~J1532$-$0039.  

\section{PHL 1811 and the High-z QSOs}

The {\it HST} STIS observation of PHL 1811 was performed 2001 December
3 and the two {\it Chandra} observations were performed 2001 December 5
and 17.  Figure~1 shows the spectral energy
distribution of PHL 1811.  The steep $\alpha_{ox}$ show that PHL 1811
was X-ray weak compared with typical quasar of its luminosity.  It
appears to be intrinsically X-ray weak, rather than absorbed, because
it is variable between two observations twelve days apart, a fact that
rules out the possibility that the central engine is absorbed and we
observe scattered X-rays.  Furthermore, the spectral index is nominal for
a NLS1 ($\Gamma\sim 2.25$), rather than flat as it might be if it were
absorbed.  The details of these observations will be presented in
Leighly, Halpern \& Jenkins, in preparation.

The three high-z lineless quasars that we considered were
SDSS~J1532$-$0039 (z=4.62; Fan et al.\ 1999) and two reported by
Anderson et al.\ 2000 (SDSS J1302$+$0030 and SDSS~J1442$+$0110; z=4.5 and
z=4.56, respectively).  Since the time that these were reported,
apparently a number of others have been discovered (see the
contribution by Collinge, these proceedings).  Figure 2 shows a
comparison of the PHL 1811 UV spectrum, shifted to z=4.5, with the
SDSS spectrum of SDSS~J1302$+$0039.  The Keck spectrum of
SDSS~J1532$-$0039 published by Fan et al.\ 1999 has a much better
signal-to-noise ratio.  To accentuate the contrast between these
objects and typical quasars, we also show the Francis et al.\ (1991)
LBQS quasar compliation spectrum shifted to z=4.77, and the SDSS
spectrum of an arbitrarily-chosen high-redshift quasar.

\begin{figure}
\plotfiddle{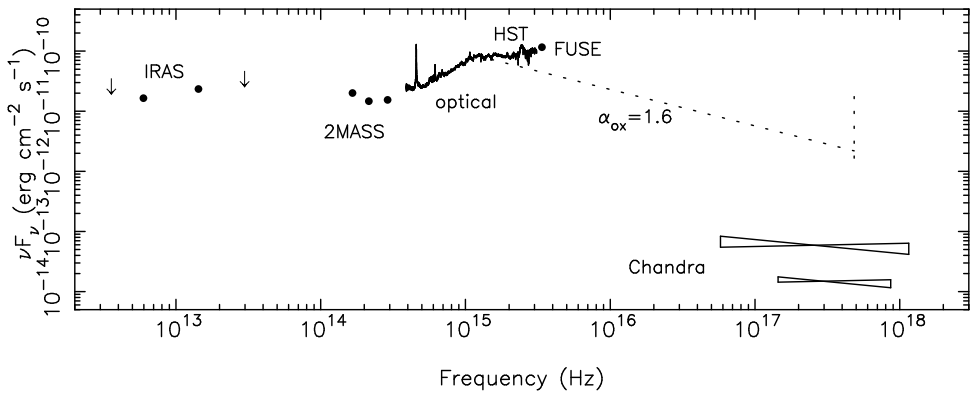}{2.0in}{0}{75}{75}{-230}{-210}
\caption{The spectral energy distribution of PHL 1811.   The observed
$\alpha_{ox}$ are $1.9$ and $2.1$; the 
location of $\alpha_{ox}=1.6$, the typical value for a quasar of this
luminosity, is shown.}
\end{figure}

\begin{figure}
\plotfiddle{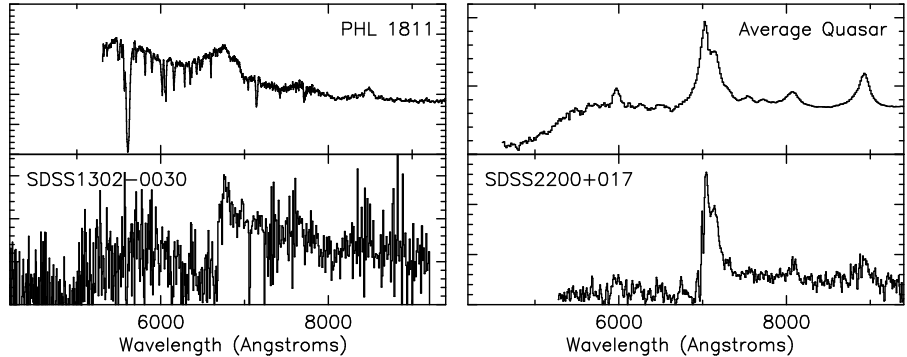}{2.0in}{0}{75}{75}{-230}{-210}
\caption{{\it Left:} The top panel shows PHL 1811, shifted to a
redshift of 4.5.  The bottom panel shows the SDSS spectrum of one of
the high-z lineless quasars.  The similarity can be seen, if one takes
in to account the difference in signal-to-noise ratio and the presence
of Ly$\alpha$ absorption in the high redshift quasar. {\it Right:} For
comparison, the top panel shows the LBQS quasar compilation spectrum
from Francis et al.\ 1991 shifted to z=4.77, and the bottom panel
shows arbitrarily chosen SDSS high redshift quasar SDSS~J2200$+$017.}
\end{figure}

\section{Discussion and Conclusions}

In this contribution, we discuss the similarity between the UV
spectrum of PHL 1811 and the spectra of several high-z lineless
quasars discovered in the SDSS, and propose that the physical cause of
their spectra is the same.  At the same time, we acknowledge that the SDSS
is quite likely to find lineless quasars whose spectra have a
different origin; for example, they may be Bl~Lac objects.

If it is true that these spectra are a result of the same phenomenon,
what are the potential implications?  Because of their steep X-ray
spectra and high amplitude X-ray variability, it has been proposed
that NLS1s are characterized by a high accretion rate (e.g., Leighly
1999 and references therein).  Since emission-line widths should
depend on the black hole mass (Laor 2000), luminous NLS1s should have
exceptionally high accretion rates compared with their Eddington
value.  It is important to understand the accretion rate in high
redshift objects, as luminous quasars have large black holes, and in
order to grow large in the short amount of time implied by the high
redshift, they should be accreting at a rapid rate.  Naturally, people
are searching for a connection between high redshift quasars and NLS1s
(e.g., Mathur 2000), although the evidence so far is mixed (Constantin
et al.\ 2002).  Based on the similarities between PHL 1811 and the
high-z lineless quasars, we suggest that they are the early Universe
counterparts of luminous Narrow-line Seyfert 1 galaxies.

What causes the weak emission lines?  In PHL 1811, the
higher-ionization emission lines are probably weak because the
continuum is very soft overall.  Such a soft continuum can produce
copious H$^0$ and Fe$^+$ ions, but few higher ionization species.
Also, the equivalent widths of the lines will appear small against the
strong, blue optical/UV continuum.


\begin{references}
\reference Anderson, S.\ F., et al.\ 2001, \aj, 122, 503
\reference Constantin, A., \& Shields, J.\ C. 2003, \pasp, 115, 592
\reference Fan, X., et al.\ 1999, \apj, 526, L57
\reference Francis,  P.\ J., Hewett, P.\ C., Foltz, C.\ B., Chaffee,
F.\ H., Weymann, R.\ J., \& Morris, S.\ L. 1991 \apj, 373, 465
\reference Laor, A. 2000, New Astr.\ Rev., 44, 503
\reference Leighly, K.\ M. 1999, \apjs, 125, 297 \& 317
\reference Leighly, K.\ M., Halpern, J.\ P., Helfand, D.\ J., Becker,
R.\ H., \& Impey, C.\ D. 2001, \aj, 121, 2889
\reference Mathur, S. 2000, \mnras, 314, 17p
\reference Vignali, C., Brandt, W.\ N., Fan, X., Gunn, J.\ E., Kaspi,
S., Schneider, D.\ P., \& Strauss, M.\ A. 2001, \aj, 122, 2143
\reference Wilkes, B.\ J., Tananbaum, H., Worrall, D.\ M., Avni, Y.,
Oey, M.\ S., \& Flanagan, J. 1994, \apjs, 92, 53
\end{references}
\end{document}